\documentclass[twocolumn,showpacs,amsmath,amssymb,pre]{revtex4}

\usepackage{graphicx}% Include figure files
\usepackage{dcolumn}% Align table columns on decimal point
\usepackage{bm}% bold math

\newcommand{\nquad}{ \!\!\!\!}

\newcommand{\expc}[1]{{\left<#1\right>}}
\newcommand{\re}{\mathop{\rm Re}} \newcommand{\Z}{{\mathcal Z}}
\newcommand{\pp}{{\mathcal P}}
\newcommand{\HH}{\mathsf{H}}
\newcommand{\ZZ}{\mathsf{Z}}
\newcommand{\mf}{{\hbox{\scriptsize mf}}}

\begin{document}

\title{Solution of a statistical mechanics model for pulse formation in lasers}
\author{Omri Gat}
\email{omri@physics.technion.ac.il}
\author{Ariel Gordon}
\email{gariel@tx.technion.ac.il}
\author{Baruch Fischer}%
\email{fischer@ee.technion.ac.il}
\affiliation{%
Department of Electrical Engineering, Technion, Haifa 32000,
Israel
}%

\date{\today}

\begin{abstract}
We present a rigorous statistical-mechanics theory of nonlinear
many mode laser systems. An important example is the passively
mode-locked laser that promotes pulse operation when a saturable
absorber is placed in the cavity. It was shown by Gordon and
Fischer \cite{ourPRL} that pulse formation is a first-order phase
transition of spontaneous ordering of modes in an effective
``thermodynamic" system, in which intracavity noise level is the
effective temperature. In this paper we present a rigorous
solution of a model of passive mode locking. We show that the
thermodynamics depends on a single parameter, and calculate
exactly the mode-locking point. We find the phase diagram and
calculate statistical quantities, including the dependence of the
intracavity power on the gain saturation function, and finite size
corrections near the transition point. We show that the
thermodynamics is independent of the gain saturation mechanism and
that it is correctly reproduced by a mean field calculation. The
outcome is a new solvable statistical mechanics system with an
unstable self-interaction accompanied by a natural global power
constraint, and an exact description of an important many mode
laser system.
\end{abstract}

\maketitle

\noindent

\section{Introduction}
Lasers can produce light in continuous wave (cw) or pulsed
manners. A special pulsed operation is mode-locking, found shortly
after the laser discovery in the early 1960's \cite{PML}. Since
then, mode-locked lasers became a leading way to produce ultra
short pulses reaching today a few femtoseconds, or about two
light-wave cycles. In a mode-locked operation, many axial modes in
a broad frequency bandwidth are phase locked and thus provide one
or multiple pulses in the cavity, giving at the output a light
pulse train. The understanding of the conditions under which a
laser operates in pulsed regime rather than in continuous wave
regime is a question of great interest, both theoretical and
practical. This question has been addressed in various studies,
being referred to as ``the second threshold" (the first one being
the lasing itself) in the earlier years \cite{Haken,
haken_threshold1,new}, and recently, in the context of laser with
a saturable absorber, as the ``self starting" problem
\cite{Krausz, threshold2, threshold3,HausConfirm, KrauszAnswer,
KrauszConfirm, threshold4, MEstability}.

Formation of pulses in lasers relies on the interaction between
axial modes. Such an interaction can be provided either by
rendering the system time dependent (modulating) or by a suitable
nonlinearity in the dynamics of the system. These two methods are
commonly referred to as ``active" and ``passive" mode-locking,
respectively. One type of nonlinearity known to encourage pulsed
operation is saturable absorption. The light transmissivity
through a (fast) saturable absorber is an {\em increasing}
function of the (instantaneous) input intensity. The saturable
absorber destabilizes the laser operation into configurations
where most of the power is concentrated in short pulses. In the
frequency or mode domain the saturable absorber induces a
nonlinear four-wave-mixing interaction between the modes, as does
the Kerr effect, with the difference that it is dissipative rather
than dispersive.

The dynamics of a laser is always subject to noise. Beside the
usual noise sources present in every physical system, there is the
inevitable fundamental noise of spontaneous emission. This noise
is inherent in lasers, since it always accompanies coherent
amplification, on which lasers rely. Therefore a model of a laser
that does not take noise properly into account risks missing key
features in the physics of a laser system.

The majority of laser theories treat noise as a perturbation, if
at all, expecting it to manifest itself as fluctuations in the
laser output. However, this approach greatly underestimates the
effect of noise. It has been recently shown \cite{ourOL} that even
very weak noise (compared to the intracavity power of the laser)
is sufficient, for example, to destabilize a passively mode-locked
laser, revealing a dramatic nonperturbative effect of the noise.

We have recently developed \cite{ourPRL,GFOC} a new approach for
the many interacting mode system, with specific emphasis on
aspects of pulse formation in mode-locked lasers. We established
an analogy between the behavior of the electromagnetic field (the
mode system in a laser) in the presence of noise and equilibrium
statistical mechanics, and applied the powerful tools of
statistical mechanics to the problem of mode-locking. In
particular it was found that the entropy associated with the noise
is an essential ingredient in the theory of mode locking. This
approach gave an inherent explanation for many experimental
phenomena of mode-locked lasers, especially the existence of a
threshold and the abruptness of formation of pulses. Passive mode
locking was identified with a first-order phase transition in the
model statistical mechanics system. Many other theoretical and
experimental features \cite{CLEO} were found, among them,
hysteresis, super-heating and super-cooling, successive formation
of multiple pulses in the cavity and more.

The results of Ref.~\cite{ourPRL} were based on a mean field
analysis of the mode interaction induced by the saturable absorber
and on numerical simulations. These indicated a first order phase
transition when the effective temperature, {\it i.e.} the noise
power, or alternatively the intracavity power, is varied. The
ordered phase corresponds to a mode locked configuration, while
the disordered phase to continuous wave operation.

In this paper we show that the statistical-mechanics treatment of
mode locking suggested in \cite{ourPRL} can be concluded with an
{exact solution}. Using a recursion relation for the partition
function, we are able to develop a systematic expansion of the
free energy in decreasing powers of the number of degrees of
freedom. The first two terms in this expansion are calculated
explicitly, allowing us to show that there is indeed a single
first order phase transition associated with passive mode locking.
All the relevant thermodynamic quantities are obtained, yielding a
complete description of mode locking in the context of the present
model. We also present an improved mean field analysis, which
turns out to be exact in the thermodynamic limit.

We show that the thermodynamics depends on a {\em single}
parameter, the strength of the saturable absorption multiplied by
the intracavity power squared, and divided by the noise strength,
and the transition occurs when this parameter crosses an explicit
threshold value.

In addition, we demonstrate the universality of the results with
respect to the gain saturation mechanism (as long as it is slow).
It is well known that gain saturation, that is, the inability of
the laser amplifier to amplify the light indefinitely, has a
crucial role in understanding the dynamics and stability of lasers
\cite{Siegman} and in particular mode-locked lasers
\cite{MenyukPMLstability}. While in Ref.~\cite{ourPRL} gain
saturation was incorporated by introducing a constraint of
strictly constant intracavity power, in this paper we present a
solution with general (slow) gain saturation mechanism. The
relation between the two approaches is similar to that of the
canonical and grand-canonical ensembles in textbook statistical
mechanics, and the saturable gain acts in manner analogous to the
chemical potential. The universality property follows immediately
from the thermodynamic equivalence of ensembles. Futhermore, for a
given gain saturation function we are able to calculate the
dependence of the intracavity power on the external parameters,
and in particular its jump at the mode-locking transition, an
experimentally-measurable quantity.

\begin{figure*}[ht]
\hbox{\includegraphics[width=8cm]{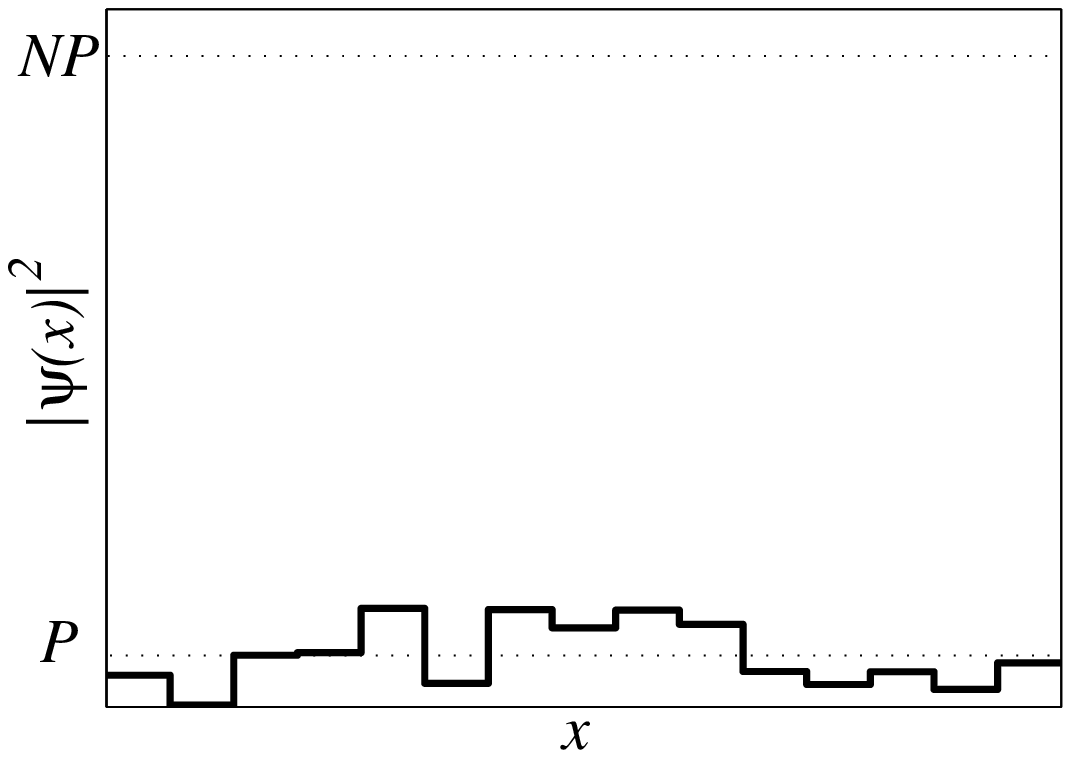}
\hskip1cm\includegraphics[width=8cm]{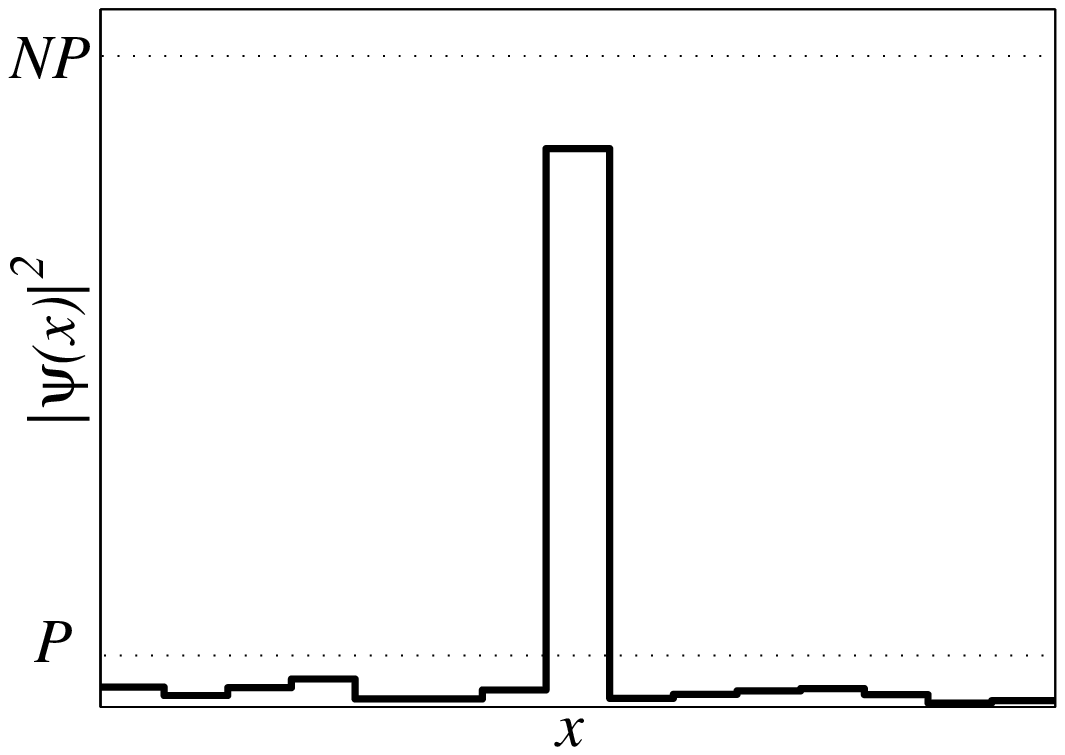}} \vskip3mm
\hbox{\hskip8mm\includegraphics[width=7cm]{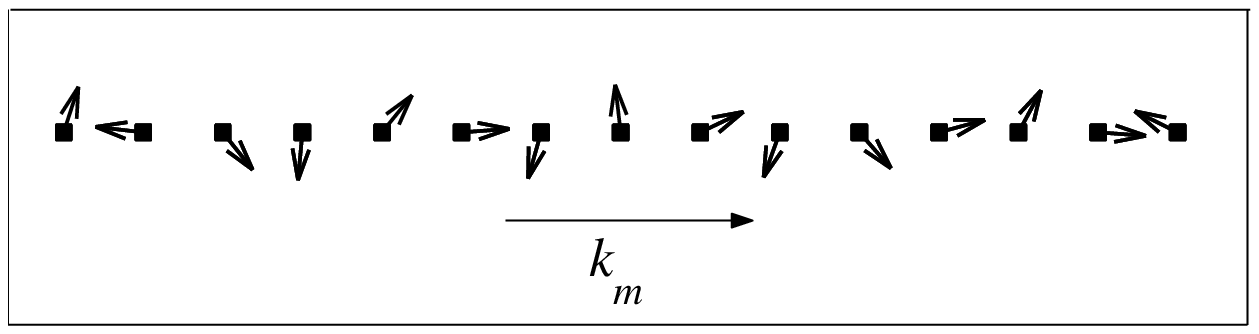}
\hskip2cm\includegraphics[width=7cm]{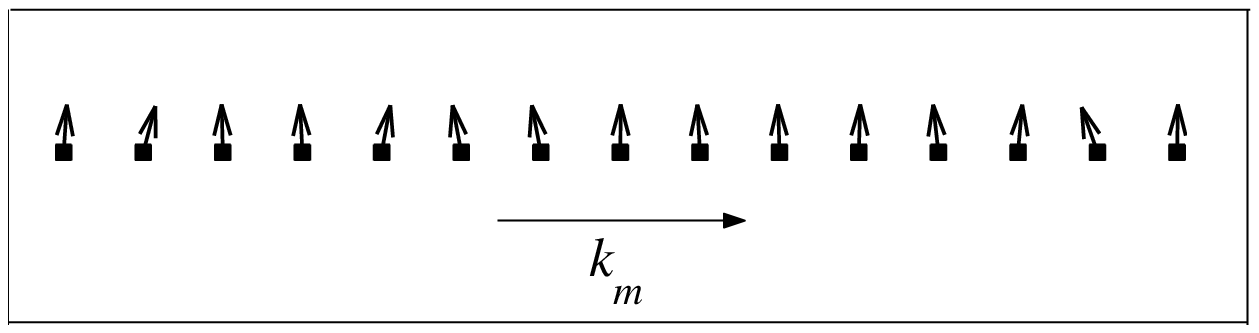}}
\caption{\label{fig:picture} The two ``thermodynamic phases'': A
typical non-mode-locked configuration is shown on the left column
of the figure and a mode-locked one on the right. The real space
configurations of the coarse grained model are shown on top row,
see Eqs. (\protect\ref{eq:z-rigid-sq-psi},\ref{eq:h-sq}) for
definitions. The Fourier modes $a_m$, defined in Eq.
(\ref{eq:fourier}) at discrete wavenumbers $k_m$, are displayed on
the bottom row as a set of phasors, each arrow representing the
complex value of a Fourier mode.}
\end{figure*}

As a statistical mechanics problem, the model can be likened to a
gas of (complex) spins, with a $|\psi|^4$ self interaction, and a
global constraint of total amplitude, quite similar to the
constraint imposed on spins in the Berlin-Kac spherical
model~\cite{berlin-kac}. Normally the statistical mechanics of
such systems leads to simple equipartition. Here, however, the
energy of self interaction term is {\em negative}, and at small
enough temperature, or high enough power the instability stemming
from the self-interaction drives the system into a pulsed state,
where most of the power resides in a single spin. In Fourier (mode
or wavenumber $k$) representation the model is equivalent to a
classical complex spin chain with a special nonlocal interaction
that drives the mode locking transition, which in this
representation is a standard ordering transition.
Fig.~\ref{fig:picture} shows the difference between typical
mode-locked and non-mode-locked configurations in Fourier space,
and in real space in the context of the coarse grained model
discussed below.

While in this work the quartic self-interaction is attributed to
saturable absorption, the model can be put into a broader
context. The quartic interaction is of general interest being the
lowest-order nonlinearity which is translation, inversion and
rotation invariant and local. This interaction itself has been
very extensively studied, but we are not aware of previous
statistical-mechanics studies of its interplay with a non-local
power constraint. In some previous well-known
statistical-mechanics studies of lasers \cite{Haken}, this
non-local constraint did not exist in the model, which may explain
why the mode-locking noise-induced threshold behavior has not
been previously found. Since a global limitation of power is
common in physical systems, our model may be an important
prototype in nonlinear optics and elsewhere.

\section{Models of passive mode locking}
\subsection{The coarse-graining approximation}
The propagation of the wave packet in a laser cavity is a
non-equilibrium process: A stationary condition is characterized
by constant pumping and dissipation of energy. Noise, due to
spontaneous emission is also present. The temporal evolution of
the wavepacket envelope $\psi$ is given schematically by
\begin{equation}\label{eq:psi}
\partial_t\psi(x,t)=G[\psi](x,t)+\eta(x,t)\ ,
\end{equation}
where $G$ is the net gain, and $\eta$ is a random driving. $\psi$
is defined on the interval $0\le x\le L$ with periodic boundary
conditions.

In the context of passive mode-locking, the net gain functional
has three essential components: $G=G_{\hbox{\scriptsize
gain}}+G_{\hbox{\scriptsize sa}} +G_{\hbox{\scriptsize sp}}$. The
net pumping of energy by stimulated emission is modelled by
\begin{equation}
G_{\hbox{\scriptsize gain}}(x,t)=
g\!\!\left(\pp\right)\psi(x,t)\ ,
\end{equation}
where $\pp=\frac1L\int_0^L|\psi(x,t)|^2\,dx$ is the total power
in the cavity. The {\em saturable gain} function $g$ is
monotonically decreasing with positive values for small $\pp$ and
negative for large $\pp$ where various losses overcome the power
supplied by the amplifier.

A necessary ingredient for passive mode locking is saturable
absorption, wherein the dissipation losses {\em decrease} as the
power increases. Unlike the saturable amplifier, the response of
the saturable absorber is fast, so it depends on the
instantaneous power $|\psi|^2$ rather than on the total power
$\pp$. We choose the specific form of saturable absorption
\begin{equation}\label{eq:satabs}
G_{\hbox{\scriptsize sa}}(x,t)=\gamma_s|\psi(x,t)|^2\psi(x,t)\ ,
\end{equation}
with $\gamma_s>0$, valid for $|\psi|^2$ not too large
\cite{HausReview}. Note that since losses are already taken into
account in $G_{\hbox{\scriptsize gain}}$, the saturable absorption
term describes an additional ``gain'' term which is absent when
$\psi=0$.

The third part models the spectral dependence of the amplifier
that has a characteristic frequency, which we assume lies at the
center of the wavepacket. Define the Fourier expansion of the
wavepacket
\begin{equation}\label{eq:transform}
\psi(x,t)=\sum_{m=-\infty}^\infty b_m(t) e^{\frac{2\pi imx}{L}}
\ . \end{equation}
Near the resonance, the pumping
efficiency falls down quadratically in the spectral
distance.
The net gain for modes off the center of the band
is therefore reduced by
\begin{equation}
G_{\hbox{\scriptsize sp}}(m,t)=
-\gamma_g \Big(\frac{2\pi m}{L}\Big)^2 b_m(t)\ ,
\end{equation}
where $\gamma_g$ is positive.

The resulting gain functional can be expressed as the functional
derivative $G(x,t)=-\delta\HH[\psi]/\delta\psi^*(x)$ of
\begin{eqnarray}\label{eq:h-soft-par}
{\HH}[\psi]&=&\int_0^Ldx\,(-\frac{\gamma_s}{2}|\psi(x)|^4+\gamma_g|\psi'(x)|^2)
\cr &+&LU\!\Big(\frac1L\int_0^L|\psi(x,t)|^2\,dx\Big)\ ,
\end{eqnarray}
with the definition $U'(\pp)=-g(\pp)$.

The random part $\eta$ of Eq.~(\ref{eq:psi}) will be modelled by
(complex) gaussian uncorrelated noise with a power of $2LT$
$$\expc{\eta^*(x,t)\eta(x',t')}=2TL\delta(x-x')\delta(t-t'),$$
$\expc{\cdot}$ stands for an ensemble average. This is appropriate
for spontaneous emission. We arrive thus at the explicit equation
of motion for $\psi$
\begin{equation}\label{eq:psi-h}
\partial_t\psi(x,t)=-\frac{\delta{\HH}[\psi]}{\delta\psi^*(x)}+\eta(x,t)\ ,
\end{equation}

It is well known~\cite{ourPRL,risken} that the invariant measure of gradient flow with
additive white noise, such as Eq.~(\ref{eq:psi-h}) is
\begin{equation}
\rho[\psi]=\ZZ^{-1}e^{-{\HH}[\psi]/(LT)}\ ,
\end{equation}
where the partition function is
\begin{equation}\label{eq:z-soft-par}
\ZZ=\int[d\psi][d\psi^*]e^{-{\HH}[\psi]/(LT)}\ .
\end{equation}
The study of passive mode locking in our model has been reduced
to the analysis the statistical mechanics system described by
$\ZZ$.

The model which is summarized by
Eqs.~(\ref{eq:h-soft-par},\ref{eq:z-soft-par}) does not include
refractive effects such as dispersion and the Kerr nonlinearity
(imaginary terms), which are important in most laser systems.
Nevertheless, as pointed out in a previous work \cite{GFOC},
$\rho$ serves as the invariant measure when dispersive effects are
included, provided that a certain integrability condition holds.
Ref.~\cite{GFOC} also studied in some detail the consequences of
the failure of the integrability condition to hold. The purpose of
the present work is different: By simplifying the model even
further, we obtain a system which captures all the essential
qualitative features of the passive mode locking transition in a
transparent and intuitive manner.

The spectral filtering term $\gamma_g|\psi'(x)|^2$ in $\HH$
introduces correlations between the electric field in neighboring
positions. It counteracts the tendency of the saturable absorption
to concentrate the power in a thinner and thinner interval by
introducing a length scale over which the electric field is
correlated. This suggests a coarse graining approximation, in
which the electric field is assumed to be constant on an interval,
and the interaction between adjacent intervals is neglected. The
function $\psi$ is then represented on the interval by a single
degree of freedom (see Fig.~\ref{fig:picture}).

Taking $N$ such intervals we find that
\begin{equation}\label{eq:h-soft-sq}
\tilde H_N[\psi]= -\frac{\gamma_s L}{2N}
\sum_{n=1}^N|\psi_n|^4-LU\!\!\left(\pp\right)
\end{equation}
where now
\begin{equation} \pp=\frac 1N\sum_{n}|\psi_n|^2
\end{equation}
 and
\begin{equation}\label{eq:z-soft-sq}
\tilde Z_N=\int\prod_n \frac{d\psi_nd\psi^*_n}{2\pi} \,
e^{{-\tilde H_N}[\psi]/(LT)}\ .
\end{equation}
The statistical mechanics problem defined by Eqs.
(\ref{eq:h-soft-sq}-\ref{eq:z-soft-sq}) is the main object of
study in this paper. At zero temperature, which corresponds to
noiseless dynamics, the quartic term in $\tilde H$ pushes all the
available power into a single degree of freedom (whose identity
depends on the initial conditions). This is the mode-locked state.
In the opposite situation at high $T$ or $\gamma_s=0$ the power
is randomly distributed, and mode-locking is absent.

\subsection{The thermodynamic limit}
Our analysis relies crucially on $N$ being very large. In the way
we stated the problem, $N$ is the ratio between the laser cavity
length and the width of a pulse. In short-pulse laser this
parameter is large by definition, with values anywhere in the
interval $10^2$--$10^9$. Taking $N\to\infty$ and expanding in
$1/N$ looks therefore promising.

It is not difficult to see, however, that the statistical
mechanics system in Eqs.~(\ref{eq:h-soft-sq}-\ref{eq:z-soft-sq})
has no well defined thermodynamic-limit behavior as $N\to\infty$.
For example, the $T=0$ value of the intracavity power $\pp$ is the
minimum of
\begin{equation}\label{eq:H0P}
-\frac 1 2 \gamma_s NL \pp^2 - LU(\pp)
\end{equation}
with respect to $\pp$, which diverges with $N$. As usual in such
cases, this means that we should ``renormalize'' the parameters of
the Hamiltonian, i.e. let $\gamma_s$, $T$ and the function $U$
depend on $N$ in such a way that a well-defined thermodynamic
limit is obtained. A comparison with a concrete example or
experiment requires a preliminary translation of the renormalized
parameters to the actual parameters.

However, a more fundamental difficulty remains: Due to the nature
of the mode-locking transition, the ratio of the peak power, the
maximal value of $|\psi_n|^2$, to the intracavity power diverges
like $N$ when $N$ is large. This means that only one of these
quantities can achieve a well-defined thermodynamic limit, while
the other quantity has to be rescaled. For example one can choose
parameters such that the peak power and $N\pp$ achieve finite
thermodynamic limiting values. This is convenient when a more
general form of saturable absorber is assumed~\cite{HausReview}.

In the present paper (as well as in Ref. \cite{ourPRL}), because
the nonlinearity is a homogeneous function (of degree 4) it is
more convenient to choose parameters such that the total power
$\pp$ reaches a well defined limit $P$, and the peak power
diverges linearly in $N$. When this is used in
Eq.~(\ref{eq:h-soft-sq}) the quartic term is seen to scale
linearly in $N$. We choose the saturable gain function such that
it scales in the same way,
\begin{equation}\label{eq:u(P)} U_N(\pp)\equiv NT u(\pp) \end{equation}
for some fixed function $u$ ($T$ is added to the definition for
later convenience). In this scheme $T$ and $\gamma_s$ need not be
renormalized.

Since the mode-locked configuration is characterized by peak power
proportional to $N$, a natural order parameter is
\begin{equation}\label{eq:M}
M=\Big(\frac 1{N^2}\sum_{n}\expc{|\psi_n|^4}\Big)^{1/4}\ ,
\end{equation}
where $\expc{}$ stands for expectation with respect to the
invariant measure. The peak power in the non-mode-locked case is
$O(1)$ in the thermodynamic limit, so $M=0$ there.

\subsection{Fixed power ensemble}
Since the quartic term in the Hamiltonian $\tilde H$ is unbounded
from below, the gain saturation term $U(\pp)$ is essential to
ensure stability, preventing the system from cascading into states
with arbitrarily low $\tilde H$. This reflects the well known fact
that lasers owe their stability to gain saturation
\cite{Siegman,MenyukPMLstability}. This is analogous to the role
of the chemical potential in the grand-canonical ensemble in
standard statistical mechanics, which limits the number of
particles in the system.

An alternative approach is to suppose that the intracavity power
$\pp$ has a \emph{fixed} value $P$, whose analogue in textbook
statistical mechanics is the canonical ensemble where the number
of particles is fixed. This is the scheme used in
Ref.~\cite{ourPRL}. One of the main results of the present work is
an equivalence of ensembles. The thermodynamics obtained in the
fixed- and variable-power ensembles are equivalent. The
variable-power ensemble has to be used if $\expc{\pp}$ is not
known. These issues are discussed in sec.~\ref{sec:power}.

In the fixed power ensemble there is no need to include the gain
saturation term, and the partition function is defined by
\begin{equation}\label{eq:z-rigid-sq-psi}
Z_N(\gamma_s,T,P)=\int\prod_n\frac{d\psi_nd\psi^*_n}{2\pi}\,
e^{\frac{-H_N[\psi]}{T}}\delta(\pp[\psi]-P)\ ,
\end{equation}
with the reduced Hamiltonian
\begin{equation}\label{eq:h-sq}
H_N[\psi]=-\frac{\gamma_s}{2N} \sum_{n=1}^N|\psi_n|^4\ .
\end{equation}
The change of variables $y_n=|\psi_n|^2$ in
Eq.~(\ref{eq:z-rigid-sq-psi}) leads to the simpler form
\begin{equation}\label{eq:z-rigid-sq}
Z_N(\gamma_s,T,P)=\int\prod_ndy_n\,e^{\frac{\gamma_s}{2NT}
\sum_{n}y_n^2} \delta\left(\frac1N \sum_{n=1}^N\!y_n-P\right)\ ,
\end{equation}
where the $y_n$ integrations are from $0$ to $\infty$. Another
change of variables $y_n\to Py_n$ leads to the useful scaling
relation
\begin{equation}\label{eq:z-scaling}
Z_N(\gamma_s,T,P)=P^{N-1}\Z_N(\gamma_sP^2/T)
\end{equation}
where $$\Z_N(\gamma)\equiv Z_N(\gamma,1,1)\      .$$ An important
conclusion has already been reached: The thermodynamics depends
on the single parameter $\gamma\equiv\gamma_sP^2/T$.
Eq.~(\ref{eq:z-scaling}) proves this in the fixed power scheme,
while the equivalence of ensembles extends this to the general
case.

In Sec.~\ref{sec:sol} we solve the statistical mechanics problem
in the fixed-power ensemble by developing an asymptotic expansion
for $\Z_N$ for large $N$, which is valid and uniform for all
$\gamma$.

\section{Mean field theory}\label{sec:mf}
The problem lends itself to a mean field analysis when formulated
in Fourier (mode) space. In the Fourier representation
mode-locking manifests as ordering in the phases of the various
modes, see Fig~\ref{fig:picture}. The discrete Fourier transform
of the Hamiltonian (\ref{eq:h-sq}) is
\begin{equation}\label{eq:h-sq-modes}
H[a]=-\frac{\gamma_s}{2}\sum_{m_1-m_2+m_3-m_4=p N}
a_{m_1}^*a_{m_2}a_{m_3}^*a_{m_4}
\end{equation}
where
\begin{equation}\label{eq:fourier}
\psi_n=\sum_{m=1}^Na_me^{2\pi im\frac nN}
\end{equation}
and $p$ is an integer (actually it is easy to see that only
$p=-1,0,1$ are relevant). $\pp$ is now given by
\begin{equation}\label{eq:rigid-modes}
\pp=\sum_{m}|a_m|^2.
\end{equation}
The main disadvantage of the Fourier space formulation is that
the nonlinear term becomes complicated and nonlocal. Mean field
theory overcomes this difficulty by assuming that the different
modes are uncorrelated and characterized by a common probability
distribution function $\rho_\mf(a)$. When the problem is
formulated in Fourier space, the mean field approximation is
quite plausible, since the interaction term involves all the
degrees of freedom.

In the mean field framework the
free energy per degree of freedom is
\begin{equation}\label{eq:mf-f}
F\equiv-\frac{\log Z}N=
-\frac{\gamma_s}{2T}N^2|\expc{a}_\mf|^4+\expc{\log\rho}_\mf\ ,
\end{equation}
where $\expc{}_\mf$ stands for expectation value
with respect to $\rho_\mf$.
Gain saturation is
included by demanding that
\begin{equation}\label{eq:mf-power}
\expc{|a|^2}_\mf=\frac P  N\ ,
\end{equation}
i.e. $\expc{\pp}_\mf=P$.

Following the standard procedure of mean field
calculations~\cite{zinn-justin}, $\rho_\mf$ is found by
minimizing the free energy subject to the
constraint~Eq.(\ref{eq:mf-power}).
A necessary condition for the minimization of $F$ is stationarity with respect to
variations of $\rho$,
\begin{eqnarray}\label{eq:mf-var}
0&=&\frac{\delta}{\delta\rho(a)}
\left(F+\lambda\left(\expc{|a|^2}-P/N\right)\right) = \\
&=&-2\re(\gamma_s/T)N^2|\expc{a}|^2\expc{a^*}a+(\log\rho(a)+1)+\lambda
|a|^2 \nonumber, \end{eqnarray}
where $\lambda$ is a Lagrange
multiplier. The solution of Eq.~(\ref{eq:mf-var}) is a gaussian
probability distribution function
\begin{equation}\label{eq:gaussian}
\rho(a)=\frac1{\pi\sigma^2}
e^{-\frac{|a-\expc{a}|^2}{\sigma^2}}\ .
\end{equation}
$\expc{a}$ and $\sigma$ are related by
Eq.~(\ref{eq:mf-power}) which implies
\begin{equation}\label{eq:variance}
\expc{|a|^2}_\mf=|\expc{a}_\mf|^2+\sigma^2=\frac P N\ .
\end{equation}
$\rho_\mf$ is therefore characterized by the single parameter
$M=\sqrt{N/P}|\expc{a}_\mf|$. From Eq.~(\ref{eq:variance}) one
can see that $0\leq|M|^2\leq 1$. When $M=0$ the phases of the
modes are completely random, which means that in real space the
power is uniformly distributed. When $M>0$ the modes are
correlated which in real space means that a macroscopic fraction
of the power resides in the variable $\psi_N$. $M$ is therefore
an order parameter, which can be shown to asymptotically coincide
with the previous definition of the order parameter
Eq.~(\ref{eq:M}).

We note that the present formulation of mean field theory is
slightly careless, in that it misses ordered states centered at
other points than $\psi_N$ in real space. The inclusion of these
configurations would require considering cases where various
$a_m$'s differ by a fixed phase. However, since the number of
missed configurations grows only linearly with $N$, the entropy is
underestimated by a subextensive factor, and the thermodynamics is
not affected.

We can define a new thermodynamic potential $f(\gamma,y)$, where
$y=M^2$ and $\gamma=\gamma_sP^2/T$, which is the free energy for a
given value $M$. Using Eqs. (\ref{eq:gaussian}-\ref{eq:variance})
in Eq.~(\ref{eq:mf-f}) it is found that
\begin{equation}\label{eq:f(m)}
f(\gamma,y)=-\left(\frac{\gamma}2 y^2+\log(1-y)\right)\ ,
\end{equation}
up to unimportant additive terms independent of $\gamma$ and $M$.
The free energy $F$ for given $\gamma$ is the global minimum of
$f(\gamma,\cdot)$, and the abscissa of the minimum, $\bar y$, is
the square of the order parameter.

\begin{figure}[ht]\label{fig:mf}
\includegraphics[width=8cm]{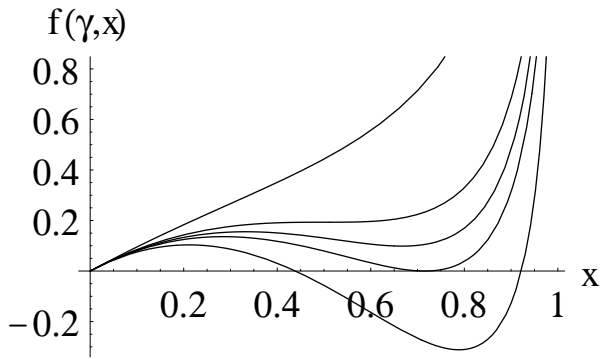}
\caption{The thermodynamic potential $f(\gamma,y)$ as a function of $y=M^2$ for
several values of $\gamma$ ranging from 2 to 6 with higher values
of $\gamma$ corresponding to lower curves. Curves with two
(local) minima correspond to systems with a metastable state.
The critical $\gamma$ ($\gamma^*$) is by definition the
one where the values of $f(\gamma,\cdot)$ at the two minima are equal,
which implies a first order phase transition.}
\end{figure}

The function $f$ of Eq.~(\ref{eq:f(m)}) has a single minimum for
$\gamma\le4$, at $y_0=0$. The vanishing of $M$ means that the
phases are not locked corresponding to the disordered, non
mode-locked phase. For $\gamma>4$ there exists an additional
(local) minimum, $y_1>0$, see Fig.~\ref{fig:mf}, which corresponds
to the mode-locked state. However, for
$4\le\gamma\le\gamma^*\simeq4.91$ $f(\gamma,y_1)>f(\gamma,y_0)$,
which means that the mode-locked state is metastable, and the true
equilibrium is still disordered, $\bar y=y_0$. At
$\gamma=\gamma^*$ the two minima exchange stability, and for all
$\gamma>\gamma^*$ the equilibrium state is mode locked, with
$\sqrt{y_1}$ as the value of the order parameter. $\gamma^*$ is
the solution of the equation
$f(\gamma,y_0)=f(\gamma,y_1(\gamma))$,
\begin{equation}
\frac{{\left( {{\sqrt{\gamma^* }+\sqrt{ \gamma^*-4 }} }
         \right) }^2}{8} = \log\frac{\sqrt {\gamma^*}(\sqrt{\gamma^* }+\sqrt{
\gamma^*-4
         })}{2}
\end{equation}

In terms of the original variables $\gamma_s$, $T$ and $P$, the
phase transition point is therefore
\begin{equation}\label{eq:trans}
\frac{\gamma_s P^2}{T}=\gamma^* \approx 4.91
\end{equation}
In order to compare it to the result in Ref. \cite{ourPRL}, one
should remember the difference in the modeling of spectral
filtering here and there. In Ref. \cite{ourPRL} we imposed
``Dirichlet" boundary conditions in the Fourier space, while here,
the coarse graining method actually induces periodic boundary
consitions in Fourier space. Since the interaction in Fourier
space is long ranged, this leads to a difference. The Hamiltonian
in Ref.~\cite{ourPRL} was identical to Eq.~(\ref{eq:h-sq-modes}),
except that $p=0$ only. The number of quartets with $k=0$ is 2/3
of their number with $p=-1,0,1$. In the mean field approximation
this would simply lead to a factor of $3/2$ in the transition
temperature:
\begin{equation}\label{eq:transprev}
\frac{\gamma_s P^2}{T}=\frac 3 2 \gamma^* \approx 7.4
\end{equation}
This is very close to the result of $\gamma_s P^2/T\approx 7.7$
obtained in Ref. \cite{ourPRL}. (We note that the mode locking
transition was specified in Ref. \cite{ourPRL} in terms of
$1/\gamma$.) The mean field theory presented here is better, since
here we do not make the ansatz of separating the modulus and angle
in $\rho_{\rm mf}$. Here we also end up with the simple analytical
expression (\ref{eq:f(m)}). Since the Hamiltonian
(\ref{eq:h-sq-modes}) is translation invariant in Fourier space,
mean field theory is more suitable for its analysis than for that
of its counterpart from Ref. \cite{ourPRL}.

In summary, it has been demonstrated in the mean field context
that the mode locking transition is a standard first order
transition, accompanied by coexistence and metastable
configurations in its neighborhood. The mean field theory
involves an uncontrolled approximation, which is hard to justify
rigorously. In the present work the justification will ultimately
follow from the real space analysis given in the next section.

Evidently, the transition condition depends, although not greatly,
on the spectral filtering scheme. Analysis of the parabolic
filtering scheme, which we do not pursue here, yields a yet
another value for the transition temperature, close to
(\ref{eq:trans}) and (\ref{eq:transprev}).

\section{Fixed power finite $N$ analysis}\label{sec:sol}
In this section we calculate an asymptotic expansion of the
partition function $Z_N(\gamma,P)$ in decreasing powers of $N$,
the number of degrees of freedom. All information pertaining to
the passive mode-locking transition is then obtainable in a
standard manner. In particular, we show that the mean field
calculations give the exact free energy.

Our starting point is a recursive version of
Eq.~(\ref{eq:z-rigid-sq}) for $Z_N$, obtained by performing only
$N-1$ of the $y$ integrations,
\begin{widetext}
\begin{equation}
Z_N(\gamma_s,T,P)= \int dy_N\,e^{\frac{\gamma_s}{2NT}y_N^2}Z_{N-1}
\Big(\frac{N-1}{N}\gamma_s,T,\frac{N-y_N}{N-1}P\Big)\ .
\end{equation}
After using the scaling relation Eq.~(\ref{eq:z-scaling}) and
making a change of the integration variable we obtain a recursive
equation for $\Z$:
\begin{equation}\label{eq:z-rec}
\Z_N(\gamma)=N\Big(\frac{N}{N-1}\Big)^{N-1} \!\!\int_0^1
dy\,e^{\frac\gamma2Ny^2}(1-y)^{N-2} \Z_{N-1}\!\!
\left(\gamma\frac{N}{N-1}(1-y)^2\right)\ .
\end{equation}
\end{widetext}
This is the fundamental equation of the real space analysis.

We will show that when $N$ is large the only significant contribution to the $y$
integration in Eq.~(\ref{eq:z-rec}) comes from the vicinity of one or two values of
$y$ which maximize the integrand, one of which is $y_0=0$. The integration on other
parts of the interval is
exponentially small in $N$ and will be neglected.
The case of a single maximizing point will be shown to correspond to invariant
measures concentrated on configurations where the amplitude of all degrees of
freedom is $O(1)$, i.e., non-mode locked, disordered configurations. This happens
for small enough $\gamma$. When there are two maximizing points the typical
configurations are such that a finite fraction of the power is concentrated in
a {\em single} degree of freedom, while the amplitude of other degrees of
freedom is again $O(1)$. These mode locked configurations arise for large enough
values of $\gamma$. We show that these are the only two possibilities.

\subsection{The disordered phase}
We first tackle the case of small $\gamma$. To this end we use the
Fourier representation of the delta function in
Eq.~(\ref{eq:z-rigid-sq}) to reexpress $\Z$ by
\begin{equation}\label{eq:zz}
\Z_N(\gamma)=\int_{-i\infty}^{i\infty}\frac{dz}{2\pi i}e^{-z}
\left(\int_0^{C_N}  dy
e^{\frac\gamma{2N}y^2+\frac{1}{N}zy}\right)^N\ .
\end{equation}
for some $C_N\geq N$. As long as $C_N\geq N$, the right-hand-side
of Eq.~(\ref{eq:zz}) is independent of $C_N$. One can now expand
the quadratic term in the exponential in a Taylor series keeping
the first two terms,  carry out the $y$ integration, and then take
$C_N\to\infty$ giving
\begin{equation}\label{eq:zn-int}
\Z_N(\gamma)\sim\int\frac{dz}{2\pi i}e^{-z} \Big(-\frac
Nz-\frac{\gamma N^2}{z^3}\Big)^N\ .
\end{equation}

The contour of integration has to be deformed so as to avoid the
singularity at $z=0$. A standard argument shows that the contour
should be moved to the left so that it crosses the real line at a
negative value. The integral (\ref{eq:zn-int}) can then be
calculated by pushing the contour through the singularity at $z=0$
and then to $\re z=\infty$. The exponential in the integrand makes
the integration at infinity vanish in the limit, leaving only
integration on a contour surrounding $z=0$ clockwise. Using
Cauchy's theorem this evaluates to
\begin{eqnarray}
\Z_N(\gamma)&\sim& N^N\oint\frac{dz}{2\pi i}
e^{-z}\sum_n{N\choose n}\frac{(\gamma N)^n}{(-z)^{2n}}=\cr&=&
\sum_n\frac{\gamma^nN^{N+n}N!}{n!(N-n)!(N+2n)!}\ ,
\end{eqnarray}
or, using Stirling's formula,
\begin{equation}\label{eq:zn-cw}
\Z_N(\gamma)\sim\frac{e^N}
{\sqrt{2\pi N}}\sum_n\frac{\gamma^n}{n!}= e^\gamma\frac{e^N}{\sqrt{2\pi N}}
\equiv\Z_N^{(0)}(\gamma)\ .
\end{equation}

$\Z_N^{(0)}$ certainly provides an asymptotic approximation of
$\Z_N(\gamma)$ as $\gamma\to 0$, but we shall show that it also
serves as the leading term of $\Z_N(\gamma)$ as $N\to\infty$ for
all $0\leq\gamma<\gamma^*$. This is achieved by showing that  the
recursive equation~(\ref{eq:z-rec})
\begin{eqnarray}\label{eq:rhs-rec-cw}
\Z_N^{(0)}(\gamma)\sim Ne\nquad &\displaystyle\int& \nquad
dy\,e^{\frac\gamma2Ny^2}(1-y)^{N-2}
\Z_{N-1}^{(0)}\left(\gamma(1-y)^2\right)\cr
=N\frac{e^N}{\sqrt{2\pi N}}&\displaystyle\int_0^\infty& dy\,
\frac{e^{\gamma(1-y)^2}}{(1-y)^2}e^{N(\frac{\gamma}{2}y^2+\log(1-y))}
\end{eqnarray}
holds for $\gamma$ in this interval. Here and below the symbol
$\,\sim\,$ stands for `asymptotic for large $N$'. Consider the
last integral. When $N\to\infty$ the integrand becomes strongly
peaked near the global minimum $\bar y(\gamma)$ of
$f(\gamma,y)=-(\frac{\gamma}{2}y^2+\log(1-y))$. The function $f$
is precisely the thermodynamic potential encountered in the
context of the mean field approximation, see Eq.~(\ref{eq:f(m)}).
As shown above (Sec.~\ref{sec:mf}), $\bar y(\gamma)=y_0\equiv0$
for $\gamma<\gamma^*$. For these values of $\gamma$ only the
neighborhood of $y=0$ has to be taken into account and the
integral in the right hand side of Eq.~(\ref{eq:rhs-rec-cw}) is
approximately
\begin{equation}
Ne^\gamma\frac{e^N}{\sqrt{2\pi N}} \int_0^1 dy\, (1-y)^{N-2}\sim
e^\gamma\frac{e^N}{\sqrt{2\pi N}}\sim \Z_{N}^{(0)},
\end{equation}
which establishes
\begin{equation}\Z_N(\gamma)\sim\Z_N^{(0)}(\gamma)
\ ,\qquad\gamma<\gamma^*.
\end{equation}

The thermodynamics now follows straightforwardly. For example, the free energy
per degree of freedom is $F(\gamma)=\frac1N\log\Z_N\sim1$, independent of $\gamma$
to leading order, and the expectation value of $|\psi|^4$ in the invariant measure is
\begin{equation}\label{eq:psi4-cw}
\expc{|\psi|^4}=2
\frac{\Z_N'(\gamma)}{\Z_N(\gamma)}\sim2\quad(\gamma<\gamma^*)\ ,
\end{equation}
also independent of $\gamma$ in the leading order. In particular,
the order parameter $M$ from Eq.~\ref{eq:M} equals zero, showing
that this is indeed a disordered configuration.

\subsection{The mode-locked phase}
We turn now to the case $\gamma>\gamma^*$, where $\bar y=y_1>0$.
We can no longer expect that $\Z_N\sim\Z_N^{(0)}$, but the mean
field calculations suggest that
\begin{equation}\label{eq:zn-ansatz}
\Z_N(\gamma)\sim A_N(\gamma)e^{-NF(\gamma)}
\end{equation}
where $F(\gamma)=f\big(\gamma,\bar y(\gamma)\big)-1$ and $A_N$ is
subexponential in $N$. The results of the previous subsection
imply that the asymptotic form~(\ref{eq:zn-ansatz}) is valid for
$\gamma<\gamma^*$, since then $\bar y=y_0$, and $F=-1$. Using
Eq.~(\ref{eq:z-rec}) for $\gamma>\gamma^*$, we presently show that
Eq.~(\ref{eq:zn-ansatz}) is valid for all $\gamma\ne\gamma^*$ and
find explicit expressions for $A_N$.

Substituting Eq.~(\ref{eq:zn-ansatz}) in Eq.~(\ref{eq:z-rec})
gives the asymptotic equation
\begin{widetext}
\begin{equation}
A_N(\gamma)e^{-NF(\gamma)}\sim N{e}\int dy\,
\frac{e^{\frac\gamma2y^2}}{1-y}A_{N-1}\Big(\gamma(1-y)^2\Big)
e^{-(N-1)\left(f(\gamma,y)+F(\gamma(1-y)^2)\right)}\ .
\label{eq:rhs-rec}\end{equation}\end{widetext} As before, the
integration is concentrated near the maximal points of the large
exponential, i.e. the minima (as a function of $y$) of
$f(\gamma,y)+F\big(\gamma(1-y)^2\big))$. Recalling the definition
of $F$ the minimization problem turns into
\begin{eqnarray}
&\phantom {=}&\min_y f(\gamma,y)+F(\gamma(1-y^2))=\cr&=&
\min_{y,\tilde w} -\gamma y^2-\log(1-y) - \gamma(1-y)^2\tilde
w^2-\log (1-\tilde w) =\cr&=&\min_{y,w} -\frac\gamma2(y^2+w^2)
+\log(1-y-w),
\end{eqnarray}
where we have put $w=\tilde w(1-y)$. It is straightforward to
check that either $y$ or $w$ must vanish at the minimum. For
$\gamma>\gamma^*$ there are two possibilities,
$y=y_1(\gamma),w=0$, and $y=0,w=y_1(\gamma)$, and the minimal
value is the same in both cases.

The conclusion is that the integration receives two main contributions, one from
the neighborhood of $y=0$, and one from the neighborhood of $y=\bar y$, which
we denote by $I_0$ and $I_1$ respectively. $I_0$ is found by approximating the
exponential near $y=0$ and evaluating prefactors at $y=0$ giving
\begin{widetext}
\begin{equation}\label{eq:rhs-rec-0}
I_0\sim\displaystyle NeA_{N-1}(\gamma)e^{-(N-1)F(\gamma)}\int_0^1
dy\, (1-y)^{N-2}e^{2(N-1)F'(\gamma)y}
\sim\displaystyle\frac{e^{F(\gamma)+1}}{1-2F'_1(\gamma)}
A_N(\gamma)e^{-NF(\gamma)}\ ;
\end{equation}
\end{widetext}
the assumption that $A_N$ is subexponential in $N$ was used to
approximate $A_{N-1}\sim A_N$.

For the calculation of $I_1$ we need to evaluate $F$ near
$\gamma(1-\bar y(\gamma))^2$. It follows from the properties of
$f$ that this is always strictly less than $\gamma^*$, where
$F\equiv-1$. Therefore
\begin{equation}\label{eq:rhs-rec-bar}
\begin{array}{rl}
I_1\sim &\displaystyle \frac{\sqrt{N}e^{\gamma(1-\bar y)^2}
}{\sqrt{2\pi}(1-\bar y)^{2}} {e^{NF(\gamma)}}
\int_{-\infty}^\infty dy\, e^{(N-1)F^{(2)}
(\gamma)\frac{(y-\bar y)^2}2}\\
\sim&\displaystyle \frac{e^{\gamma(1-\bar y)^2}}{(1-\bar y)^2
\sqrt{F^{(2)}(\gamma)}}e^{-NF(\gamma)}\ .
\end{array}\end{equation}
where $F^{(2)}(\gamma)=\partial_y^2f\big(\gamma,\bar
y(\gamma)\big)$. Combining these results with
Eq.~(\ref{eq:rhs-rec}), we get a linear equation for $A_N$ whose
solution is
\begin{equation}\label{eq:zn-ml}
A_N(\gamma)=\frac{e^{\gamma(1-\bar y)^2} } {(1-\bar
y)^2\sqrt{F^{(2)}}}
\frac{1-2F'(\gamma)}{1-2F'(\gamma)-e^{F(\gamma)+1}}\quad\gamma>\gamma^*
\ .
\end{equation}
This establishes equation Eq.~(\ref{eq:zn-ansatz}) , with explicit
values for $A_N$ for all $\gamma\ne\gamma^*$.

We find now that $F(\gamma)$ is indeed the free energy,
consistently with the mean field theory. Since
\begin{equation}\label{eq:psi4-ml}
\expc{|\psi|^4}\sim 2(\log\Z_N)'(\gamma)=
2\frac{A'(\gamma)}{A(\gamma)}-2NF'(\gamma)\ ,
\end{equation}
the order parameter $M=\big(2F'(\gamma)\big)^{1/4}$ is nonzero for
$\gamma>\gamma^*$, showing that mode locking occurs for such $\gamma$.\
Using the definition of $F$ we can calculate explicitly
\begin{equation}
-2F'(\gamma)=-\frac{d}{d\gamma}f\big(\gamma,\bar y(\gamma)\big)=
-{\partial_\gamma}f\big(\gamma,\bar y(\gamma)\big)=\bar y^2\ ,
\end{equation}
since $f$ is by definition stationary with respect to $y$ at
$\bar y$. This means that $M=\sqrt{\bar y}$, also in accordance
with the mean field calculations. It is possible to show by
calculating higher moments that $M^2$ is the power concentrated
in a {\em single} degree of freedom. This result, which in
physical terms means that mode locking results in a single pulse,
can be traced to the fact that the integral in
Eq.~(\ref{eq:rhs-rec}) receives contributions only from $y_0$ and
$y_1$.

\subsection{The transition region}
The analysis of the previous section does not apply to the case
where $\gamma$ is precisely equal to $\gamma^*$. For example,
Eq.~(\ref{eq:zn-ml}) would imply that $A_N(\gamma^*)$ is infinite,
since the derivative in $F'(\gamma)$ should be taken from below.
More importantly, the asymptotic approximation
Eq.~(\ref{eq:zn-ansatz}) is not uniform in $N$ near $\gamma^*$,
because it neglects the contribution of the metastable state. An
asymptotic approximation for $\Z$ which is valid and uniform for
all $\gamma$ is
\begin{equation}\label{eq:zn-unif}
\Z_N(\gamma)\sim e^\gamma\frac{e^N}{\sqrt{2\pi N}}+
A_N^*(\gamma)e^{-NF^*(\gamma)} \ ,\end{equation}
where $F^*(\gamma)=f\big(\gamma,y_1(\gamma)\big)$, and $A_N^*$ is given
by Eq.~(\ref{eq:zn-ml}) replacing everywhere $F$ by $F^*$ and
$\bar y$ by $y_1$. The uniform approximation is a {\em continuous}
function of $\gamma$ which reduces to the non-uniform
approximation for $|\gamma-\gamma^*|\gg1/N$.

\begin{figure}[ht]\label{fig:transition}
\includegraphics[width=8cm]{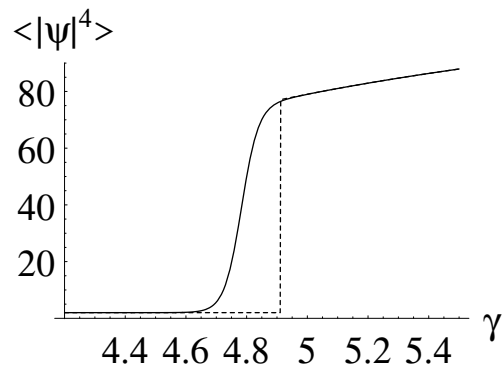}
\caption{A comparison between the value of the order parameter in
the uniform (full line) and the non-uniform (broken line)
approximation as a function of the nonlinearity parameter $\gamma$
for a system with $N=150$ degrees of freedom. The non-uniform
approximation breaks down near the transition point $\gamma^*$.}
\end{figure}

Observables in systems with a finite number of degrees of freedom
exhibit crossover behavior in the mode-locking transition, rather
than the sharp, discontinuous dependence on parameters predicted
in the thermodynamic limit. When the number of degrees of freedom
is not too large, the crossover is measurable and describable by
the uniform approximation. For example,
\begin{equation}\label{eq:psi4-unif}
\expc{|\psi|^4}\sim 2\big(a_N(\gamma)+N b_N(\gamma) y_1(\gamma)^2\big)\ ,
\end{equation}
where
\begin{equation}%\textstyle
a_N=\frac{e^{\gamma}\frac{e^N}{\sqrt{2\pi
N}}+(A_N^*)'(\gamma)e^{-NF^*(\gamma)}}
{e^{\gamma}\frac{e^N}{\sqrt{2\pi N}}+A_N^*e^{-NF^*(\gamma)}}
\end{equation}
and
\begin{equation}%\textstyle
b_N=\frac{A_N^*(\gamma)e^{-NF^*(\gamma)}}
{e^{\gamma}\frac{e^N}{\sqrt{2\pi N}}+A_N^*e^{-NF^*(\gamma)}}\ .
\end{equation}
For moderate values of $N$, there is a significant interval
in $\gamma$ below the transition where $\expc{|\psi|^4}$ is
much larger than its value in the thermodynamic limit.
A comparison between the uniform and non-uniform approximations
to $\expc{|\psi|^4}$
is shown in Fig.~\ref{fig:transition}.

\section{Thermodynamics with variable total power}
\label{sec:power} In the previous section we developed a
systematic approximation scheme for the partition function $Z$ of
the passive mode locking model as a function of the nonlinearity
strength $\gamma_s$, the fixed total intracavity power $P$ and
the temperature $T$. This allowed us to calculate the free energy
energy per degree of freedom, and we found that mode locking
occurs whenever $\gamma_s P^2/T$ is greater than a critical value
$\gamma^*$.

However, in experimental situations the intracavity power $\pp$ is
not fixed in advance. Rather it is a fluctuating quantity, whose
mean value $P$ is determined by the saturable gain function $U$
(see Eq.~(\ref{eq:h-soft-sq})). The relation between the
thermodynamics in the fixed-power ensemble analyzed above, and the
variable-power ensemble which is the subject of this section is
quite similar to the one between the canonical and grand canonical
ensembles in statistical mechanics \cite{landau-lifshitz}. In the
latter case one defines the grand potential $\Omega=\mu N-F$,
where $\mu=\partial F/\partial N$ is the chemical potential. An
equivalent thermodynamics is obtained after replacing the
extensive variable $N$, by the intensive variable $\mu$. Finite
size corrections to the thermodynamics in the two ensembles are
also related, but not equivalent.

In this section we show in a similar spirit that thermodynamics
with fixed and variable power is equivalent, and calculate the
thermodynamic limit of $P\equiv\expc{\pp}$ in the variable power
case. It is quite straightforward to generalize the calculations
and to obtain subleading terms as in Sec.~\ref{sec:sol}, but this
is not pursued here.

In the fixed power ensemble the free energy per degree of freedom
$\tilde F$ of the Hamiltonian $\tilde H_N$ which includes the
saturable gain is related to the free energy $F$ calculated in
Sec.~\ref{sec:sol} by
\begin{equation}
\tilde F(\gamma_s,T,P)=F(\gamma_sP^2/T)
+\log P+u(P)\ .\end{equation}
(Refer to Eqs.(\ref{eq:h-soft-sq},\ref{eq:u(P)}) for
the relevant definitions).
We now define the variable-power
thermodynamic potential $\Phi$ as the Legendre transform of $\tilde F$,
\begin{equation}\label{eq:Phi}
\Phi(\gamma_s,T,\mu)=\min_P \mu P-\tilde F(\gamma_s,T,P)\ .
\end{equation}
The function $u$ has to grow faster than $P^2$ to ensure that the
minimum in Eq.~(\ref{eq:Phi}) is finite. We also require that $u$
is convex to obtain a unique minimum. Otherwise $u$ is arbitrary.

The thermodynamics is obtained from $\Phi$ through the properties of the
Legendre transform, the case of interest being $\mu=0$. The mean power,
\begin{equation} P=\partial_\mu\Phi(\gamma_s,T,0)\ ,\end{equation}
can be found from the definition Eq.~(\ref{eq:Phi}) and the results of
the previous section; it is given implicitly by
\begin{equation}\label{eq:p-max}
1+Pu'(P)-\gamma \bar y(\gamma)^2=0\ ,
\end{equation}
where as before $\gamma=\gamma_sP^2/T$. The order parameter is
\begin{equation}
M^4=2T\partial_{\gamma_s}\Phi=-2T\partial_{\gamma_s}\tilde F
=-P^2F'(\gamma)=P^2\bar y(\gamma)^2\ ,
\end{equation}
which, for a given mean power $P$,
is {\em independent} of the form of $g$, and therefore also
equal to the order parameter in the
fixed power ensemble. Moreover, the
thermodynamics depends on the single parameter
$\gamma$. A special case is the
mode-locking transition point, which occurs at $\gamma=\gamma^*$
whatever the form of the saturable gain function. This universal
behavior stems
from the thermodynamic equivalence of the fixed power and variable power
ensembles.

Another interesting thermodynamic quantity which can be studied only
in the variable power framework is the susceptibility
$\chi=P'(\gamma)$ which measures the response of the intracavity power to
changes in the strength of the nonlinearity or inverse noise power.
Taking the derivative of Eq.~(\ref{eq:p-max}) shows that
\begin{equation}
\chi=\frac{\big(\gamma\bar y(\gamma)^2\big)'}{(Pu'(P))'}\ .
\end{equation}
In the non-mode locked regime $P$ is independent of $\gamma$ and
$\chi=0$; when $\bar y>0$ and mode-locking occurs,
$\bar y'(\gamma)$ is also positive and it follows from the convexity of
$u$ that the susceptibility is strictly positive in mode-locked
systems.

\begin{acknowledgments}
We are pleased to acknowledge fruitful discussions with Shmuel
Fishman. This work was supported by the Israeli Science Foundation
(ISF) founded by the Israeli Academy of Sciences.
\end{acknowledgments}


\begin{thebibliography}{}
\bibitem{ourPRL}  A. Gordon and B. Fischer,
Phys. Rev. Lett. {\bf89}, 103901, (2002)



\bibitem{PML} H.W. Mocker and R. J. Collins, Appl. Phys.
Lett. {\bf 7}, 270 (1965).

\bibitem{haken_threshold1} H. Haken and H. Ohno, Opt. Commun. {\bf
16}, 205 (1976)
\bibitem{Haken} H. Haken, ``Synergetics", 2-nd enlarged edition,
Springler-Verlag, Berlin Heidelberg New-York (1978)
\bibitem{new} G. H. C. New, Proc. IEEE {\bf 67} 380 (1979)

\bibitem {Krausz} F. Krausz, T. Brabec and Ch. Spilmann, Opt. Lett. {\bf 16} 235 (1991)

\bibitem{threshold2} H. A. Haus and E. P. Ippen,  Opt. Lett. {\bf 16}, 1331 (1991)

\bibitem{threshold3} J. Herrmann, Opt. Commun. {\bf 98} 111 (1993)


\bibitem{HausConfirm} K. Tamura, J. Jacobson, E. P. Ippen, H. A. Haus, and J. G.
Fujimoto, Opt. Lett. {\bf 18} 220 (1993)

\bibitem{KrauszAnswer} F. Krausz and T. Brabec, Opt. Lett. {\bf 18}, 888
(1993)


\bibitem{KrauszConfirm} Y.-F. Chou, J. Wang, H.-H. Liu, and N.-P. Kuo, Opt. Lett. {\bf 19} 566
(1994)

\bibitem{threshold4} C. J. Chen, P. K. A. Wai and C. R. Menyuk, Opt. Lett. {\bf 20}, 350 (1995)
\bibitem{MEstability} T. Kapitula, J. N. Kutz and B. Sandstede, J. Opt. Soc. Am. B {\bf 19}, 740, (2002)
\bibitem{ourOL} A. Gordon and B. Fischer, Opt. Lett {\bf 18} 1326
(2003).
\bibitem{GFOC} A. Gordon and B. Fischer, Opt. Comm. \textbf{223},
151 (2003).


\bibitem{CLEO} A. Gordon {\emph et al.} {\emph in Proceedings of the
Conference on Lasers and Electro-Optics, Baltimore, 2004}, session
CWM6.


\bibitem{HausReview} H. A. Haus, %``Mode-Locking of Lasers",
IEEE J. Sel. Top. Quant. {\bf 6} 1173 (2000)

\bibitem{Siegman} A. E. Siegman, ``lasers", Mill Valey, CA,
University of Science Books, 1986

\bibitem{MenyukPMLstability} C. J. Chen, P. K. Wai and C. R.
Menyuk, Opt. Lett. {\bf 19} (3) 198 (1994)

\bibitem{berlin-kac} T. H. Berlin and M. Kac, Phys. Rev. {\bf 86} 821
(1952)

\bibitem{risken} H. Risken, ``The Fokker-Planck Equation", second
edition, Springler-Verlag (1989).

\bibitem{landau-lifshitz} L. D. Landau and E. M. Lifshitz,
"Statistical Physics",3-rd ed. part 1, Oxford: Pergamon (1980).

\bibitem{zinn-justin} J. Zinn-Justin, "Quantum Field Theory
and Critical Phenomena",  3rd ed., Oxford : Clarendon Press,
(1996).





\end{thebibliography}
\end{document}